\documentclass[a4paper,fleqn]{cas-dc}

\usepackage[english]{babel}
\usepackage[numbers,sort&compress]{natbib}
\usepackage{graphicx}
\usepackage{graphics}
\usepackage{braket}
\usepackage{bbold}
\usepackage{amssymb}
\usepackage{amsmath}
\usepackage{nicefrac}
\usepackage{dcolumn}
\usepackage{bm}
\usepackage{slashed}
\usepackage{datetime}
\usepackage{mciteplus}
\usepackage{multirow}
\usepackage{bbold}
\usepackage{color}
\usepackage{xcolor}
\usepackage{float}
\usepackage[utf8]{inputenc}
\usepackage[normalem]{ulem}

\def\tsc#1{\csdef{#1}{\textsc{\lowercase{#1}}\xspace}}
\tsc{WGM}
\tsc{QE}

\newcommand{\drv}{{\rm d}}

\newcommand{\MSb}{\overline{\rm MS}}

\newcommand{\CmHENLO}{C_m^{{\rm HE}\text{-}{\rm NLO}}}
\newcommand{\DY}{\Delta Y}
\newcommand{\JPsi}{J/\psi}
\newcommand{\Yps}{\Upsilon}

\newcommand{\tref}[1]{~\ref{#1}}
\newcommand{\eref}[1]{~\eqref{#1}}

\begin{document}
\let\WriteBookmarks\relax
\def\floatpagepagefraction{1}
\def\textpagefraction{.001}

\shorttitle{The high-energy spectrum of QCD from inclusive emissions of charmed $B$-mesons}    

\shortauthors{Celiberto, Francesco Giovanni}  

\title []{\Huge The high-energy spectrum of QCD from inclusive emissions of charmed $B$-mesons}  

\author[1,2,3]{Francesco Giovanni Celiberto}[orcid=0000-0003-3299-2203]

\cormark[1]


\ead{fceliberto@ectstar.eu}


\affiliation[1]{organization={European Centre for Theoretical Studies in Nuclear Physics and Related Areas (ECT*)},
            addressline={Strada delle Tabarelle 286}, 
            city={Villazzano},
            postcode={I-38123}, 
            state={Trento},
            country={Italy}}

\affiliation[2]{organization={Fondazione Bruno Kessler (FBK)},
            addressline={Via Sommarive 18}, 
            city={Povo},
            postcode={I-38123}, 
            state={Trento},
            country={Italy}}

\affiliation[3]{organization={INFN-TIFPA Trento Institute of Fundamental Physics and Applications},
            addressline={Via Sommarive 14}, 
            city={Povo},
            postcode={I-38123}, 
            state={Trento},
            country={Italy}}




\begin{abstract}
 We investigate the high-energy behavior of strong interactions through a study on the inclusive hadroproduction of charmed $B$-mesons ($B_c$ or $B_c^*$ states) accompanied by non-charmed $b$-hadron or light-flavored jet emissions at LHC energies and kinematic configurations. By making use of the hybrid high-energy and collinear factorization, where the standard fixed-order description based on collinear parton densities and fragmentation functions is enhanced \emph{via} the Balitsky--Fadin--Kuraev--Lipatov (BFKL) resummation of energy logarithms, we perform a full next-to-leading order analysis of distributions differential in rapidities and azimuthal angles calculated by the hands of the {\tt JETHAD} multi-modular working package. The large observed transverse momenta justify the use of non-relativistic QCD next-to-leading order fragmentation functions to describe the heavy-flavored meson production mechanism. We come out with the conclusion that the study of this process can be included in forthcoming analyses at the (high-luminosity) LHC as a tool to access the QCD dynamics at high energies and to explore possible common ground between different resummation approaches.
\end{abstract}



\begin{keywords}
 high-energy resummation \sep
 QCD phenomenology \sep 
 charmed $B$-mesons \sep
 heavy flavor \sep 
\end{keywords}

\maketitle

\section{Introduction}
\label{sec:introduction}

Inclusive emissions of heavy-quark flavored particles in high-energy hadronic scatterings are widely recognized as excellent sounds to unveil the inner dynamics of fundamental interactions.
Heavy quarks are expected to couple with beyond-the-Standard-Model (BSM) objects, thus making them \emph{sentinels} in the search for signals of New Physics.
Yet they can serve as useful tools to make precision studies of strong interactions, the charm and bottom masses lying in a region where perturbative Quantum Chromodynamics (QCD) is at work.
Particular relevance in the QCD context has the in\-ve\-sti\-ga\-tion of reactions featuring \emph{quarkonium} production. The discovery of the $\JPsi$ in 1974~\cite{SLAC-SP-017:1974ind,E598:1974sol} provided us with a first evidence of the existence of heavy-quark flavors.
Although being easily studied at the experimental level, the theoretical description of quarkonium hadronization mechanisms still remains a challenge. Many models have been proposed so far, none of them being however capable of catching all the signatures coming from data. 

In order to solve the quarkonium production puzzle, an effective theory, called non-relativistic QCD (NRQCD), was built~\cite{Caswell:1985ui,Thacker:1990bm,Bodwin:1994jh}.
It relies on the assumption that to the physical quarkonium all possible Fock states contribute \emph{via} a linear superposition ($|Q \bar Q\rangle$, $|Q \bar Q g\rangle$, and so on). All these terms are organized as a double expansion in powers of the strong coupling, $\alpha_s$, and the relative velocity of the two constituent heavy quarks, $v$.
Cross sections for quarkonium emissions take the form of a sum of partonic hard factors featuring the emission of a given Fock state, each of them being multiplied by a long-distance matrix element (LDME) that embodies the non-perturbative information about the hadronization.

The advent of NRQCD allowed for stringent tests of quarkonium production mechanisms, in particular the \emph{short-distance} emission of a $(Q \bar Q)$ pair directly produced in the hard scattering. This channel dominates in the low observed transverse-momentum ($\bm{q}$) regions, since the pair is created with a relative transverse distance of order $1/|\bm{q}|$.
When $|\bm{q}|$ grows another mechanism becomes important, \emph{i.e.} the \emph{fragmentation} of a single parton followed by its inclusive decay into the observed quarkonium. According to collinear factorization, DGLAP-evolving fragmentation functions (FFs) need to be used to describe the hadronization. These perturbative FFs can be computed in NRQCD~\cite{Braaten:1993rw}.
First phenomenological analyses aimed at shedding light on the transition sector between short-distance and fragmentation mechanisms were conducted in the '90s~\cite{Cacciari:1994dr,Cacciari:1995yt}.
See~\cite{Lansberg:2019adr} for a review of progresses and challenges in quarkonium studies.

An intriguing opportunity to probe the QCD non-re\-la\-ti\-vi\-stic limit and, more in general, the heavy-flavor sector, emerges from the study of charmed $B$-mesons.
They are the only hadron species whose lowest Fock state is composed by two heavy quarks with different flavor: $|c \bar b\rangle$ for the positive-charged case, and $|\bar c b\rangle$ for the negative one.
At variance with quarkonia\footnote{Some authors extend the definition of quarkonia by including also charmed-bottomed bound states. In this work we adopt the traditional definition, under which quarkonia are mesons whose lowest Fock state contains a heavy quark and the correspondent antiquark, as $|c \bar c\rangle$ or $|b \bar b\rangle$.} these mesons cannot annihilate into gluon. This makes them quite stable states with narrow decay widths.
Moreover, since top quarks are extremely short-lived and they decay before hadronizing, charmed-bottomed systems are thought to be the final frontier for meson spectroscopy (see~\cite{Ortega:2020uvc} and references therein). The first observed state was the $B_c(^1S_0)$ by CDF Tevatron in 1998~\cite{CDF:1998ihx}. After almost twenty years, ATLAS detected a signature of the $B_c(^3S_1) \equiv B_c^*$ resonance~\cite{ATLAS:2014lga}. Further signals of $B_c$ excited states were reported by CMS~\cite{CMS:2019uhm} and LHCb~\cite{LHCb:2019bem} Collaborations.

The simultaneous presence of two different heavy-quark flavors in their leading Fock states makes charmed $B$-mesons excellent probe channels 
to reveal the deep nature of strong and weak interactions. 
Studies on direct productions as well as on indirect emissions from electroweak decays represent a very fertile ground where to unveil not only the $B_c^{(*)}$ production mechanisms, but also the properties of the decaying objects. 
With this aim, detailed analyses on $B_c^{(*)}$ emissions \emph{via} decays $W$-bosons~\cite{Qiao:2011yk,Liao:2012rh}, $Z$-bosons~\cite{Chang:1992bb,Yang:2010yg,Qiao:2011zc}, and top quarks~\cite{Qiao:1996rd,Chang:2007si} have been conducted.
By making use of symmetry-preserving approaches to hadronic bound states in the continuum, the semileptonic decay of charmed $B$-mesons to $\eta_c$ and $\JPsi$ was proposed as a benchmark channel for Standard-Model predictions~\cite{Yao:2021pyf}.

Notably, productions of heavy-quarkonium states and of charmed $B$-mesons can be excellent channels to investigate rare Higgs decays (see, \emph{e.g.},~\cite{Karyasov:2016hfm,Liao:2018nab}).
A NRQCD approach was employed in~\cite{Jiang:2015pah} to evaluate the $B_c^{(*)}$ production rate in Higgs boson decays both \emph{via} the short-distance and the fragmentation mechanism. A major outcome of that work is the fair evidence that such a reaction will be detectable at the high-luminosity LHC.

Higher-order calculations are a key ingredient to conduct precision studies of $B_c^{(*)}$ emissions. Next-to-leading order (NLO) calculations of these observables can be found, \emph{e.g.}, in~\cite{Qiao:2011zc,Sun:2010rw,Zheng:2019egj,Chen:2020dtu}.
Within the NRQCD formalism, collinear FFs for $\bar b$ and $c$ quarks fragmenting into $B_c$ and $B_c^*$ mesons were obtained with leading-order (LO)~\cite{Chang:1992bb,Braaten:1993jn,Ma:1994zt} and NLO accuracy~\cite{Zheng:2019gnb}.
This result was followed by the calculation of NLO gluon FFs depicting the $g \to B_c^{(*)}$ process, recently achieved in~\cite{Zheng:2021sdo} by means of the subtraction method~\cite{Artoisenet:2018dbs} and in~\cite{Feng:2021qjm} through an automated approach based
on sector-decomposition strategies~\cite{Binoth:2000ps}.
On the other hand, the validity of the fragmentation approximation for charmed $B$-mesons was explored in~\cite{Kolodziej:1995nv}, and the conclusion was that this channel starts to prevail on the short-distance mechanism when the meson is emitted with a transverse momentum larger than 10 GeV. A more critical result came out from~\cite{Artoisenet:2007xi}, where this lower bound was set to 80 GeV.

Inclusive productions of heavy-flavored hadrons can be used to access the QCD dynamics in its high-energy asymptotic limit. In this kinematic regime, large logarithms of the center-of-mass energy, $\sqrt{s}$, are enhanced. They appear to all orders of the perturbative expansion up to spoil the convergence of the QCD running-coupling series.
The most adequate framework to perform an all-order resummation of those large logarithms is the Balitsky--Fadin--Kuraev--Lipatov (BFKL) approach~\cite{Fadin:1975cb,Kuraev:1977fs,Balitsky:1978ic}, whose validity holds up to the leading logarithmic approximation (LL), which includes all contributions proportional to $\alpha_s^n \ln (s)^n$, and to the next-to-leading one (NLL), which accounts for all terms proportional to $\alpha_s^{n+1} \ln (s)^n$.

BFKL cross sections for hadronic reactions are high-energy factorized as a convolution of two impact factors and a process-independent Green's function, which encodes the resummation of energy logarithms~\cite{Fadin:1998py,Ciafaloni:1998gs,Fadin:1998jv,Fadin:2004zq} and it is known with NLO accuracy.
Impact factors are process-dependent, and a few of them have been calculated up to NLO. They contain collinear inputs such as incoming-hadron parton distribution functions (PDFs) and outgoing-object FFs.
An incomplete list of phenomenological probe channels for BFKL contains: the Mueller--Navelet channel~\cite{Mueller:1986ey} (see~\cite{Colferai:2010wu,Ducloue:2013hia,Ducloue:2013bva,Caporale:2014gpa,Celiberto:2015yba} for applications), the inclusive emission of light hadrons~\cite{Bolognino:2018oth,Celiberto:2020rxb}, multi-jet tags~\cite{Caporale:2016zkc}, Higgs plus jet~\cite{Celiberto:2020tmb,Celiberto:2022fgx}, forward Drell--Yan~\cite{Celiberto:2018muu} with a possible associated backward-jet detection~\cite{Golec-Biernat:2018kem}, and  heavy-flavored emissions~\cite{Boussarie:2017oae,Celiberto:2017nyx,Bolognino:2019ouc,Bolognino:2019yls,Bolognino:2021mrc,Celiberto:2021dzy,Celiberto:2021fdp}.

Major issues in attempts at precision studies of Mueller--Navelet cross sections and azimuthal correlations were highlighted in~\cite{Ducloue:2013hia,Ducloue:2013bva,Caporale:2014gpa}. More in particular, NLL corrections both to the BFKL Green's function and impact factors come out with same weight and opposite sign of genuine LL results. This makes the high-energy series unstable, its main manifestation emerging when renormalization and factorization scales are varied around the \emph{natural} energies suggested by kinematics. The presence of those instabilities was later confirmed also when light hadrons are detected~\cite{Celiberto:2020wpk}.

A first unambiguous signal of a reached stability of high-energy predictions under NLL corrections at natural scales emerged quite recently in the context of inclusive detections of heavy-light hadrons, $\Lambda_c$ baryons~\cite{Celiberto:2021dzy} and bottom-flavored hadrons~\cite{Celiberto:2021fdp}, whose lowest Fock state contains either the charm or the bottom quark. In these studies it was discovered that the typical pattern of variable-flavor-number-scheme (VFNS)~\cite{Buza:1996wv} collinear FFs depicting the parton hadronization to heavy-light hadrons at large $|\bm{q}|$ leads to a \emph{natural stabilization} of the high-energy resummation.
These stabilizing effects were soon after observed also in inclusive emissions of $\JPsi$ or $\Yps$ accompanied by light-jet tags~\cite{Celiberto:2022dyf}. That study relied on the fragmentation approximation for the quarkonium production at large $|\bm{q}|$. More in particular, a DGLAP-evolved FF set based on a NQRCD NLO input~\cite{Braaten:1993mp,Zheng:2019dfk}, called {\tt ZCW19}, was built. In view of these results, it becomes relevant the search for a possible systematic stabilization of high-energy dynamics through the production of other heavy-quark bound states that can by described \emph{via} the NRQCD fragmentation.

In this work we make use of rapidity and azimuthal-angle differential distributions for the inclusive production of $B_c^{*}$ mesons accompanied by other bottomed states (non-charmed $B$-mesons and $\Lambda_b$ baryons, comprehensively labeled $b$-hadrons or ${\cal H}_b$) or by light jets, as tools to access the high-energy spectrum of QCD within the NLL accuracy.

We believe that the analyses proposed here could serve as common basis for a win-win strategy. The hunt for the aforementioned stabilizing effects is needed to assess the feasibility of precision studies at high energies, which in turn can be employed to investigate the production mechanisms of charged $B$-mesons, and to possibly validate the use of the NRQCD fragmentation.

\section{Theoretical framework}
\label{sec:theory}

We investigate the following hadroproduction reaction
\begin{eqnarray}
\label{process}
 {\rm proton}(p_a) + {\rm proton}(p_b) \to B_c^{(*)}(q_1, y_1) + {\cal X} + {\cal P}(q_2, y_2) \;,
\end{eqnarray}
where a charmed $B$-meson is inclusively emitted in association with a $b$-hadron or a light-flavored jet, ${\cal P} = \{ {\cal H}_b, {\rm jet} \}$, and together with an undetected gluon-radiation system, ${\cal X}$. The two particles are emitted with transverse momenta $|\bm{q}_{1,2}| \gg \Lambda_{\rm QCD}$, and a large rapidity interval, $ \DY \equiv y_1 - y_2$.
The $q_{1,2}$ four-momenta are decomposed \emph{à la} Sudakov on the basis of the colliding-proton momenta, $p_{a,b}$, thus having
\begin{eqnarray}
\label{sudakov}
q_{1,2} = x_{1,2} p_{a,b} + \frac{\bm{q}_{1,2}^2}{x_{1,2} s}p_{b,a} + q_{1,2\perp} \ , \quad
q_{1,2\perp}^2 = - \bm{q}_{1,2}^2 \;.
\end{eqnarray}
In the center-of-mass system rapidities are connected to corresponding final-state longitudinal momentum fractions, $x_{1,2}$, \emph{via} the relation
$y_{1,2} = \pm \ln (x_{1,2} \sqrt{s} / \bm{q}_{1,2}^2)$.

\subsection{NLO/NLL cross section}
\label{sec:cross_section}

The cross section differential in the rapidity distance, $\DY$, and in the difference between the azimuthal angles of the detected particles, $\varphi \equiv \varphi_1 - \varphi_2 - \pi$, can be cast in the form of a Fourier sum of azimuthal coefficients, $C_{m \ge 0}$
\begin{eqnarray}
 \label{dsigma_Fourier}
 \frac{\drv \sigma}{\drv \varphi \, \drv \DY} =
 \frac{1}{2\pi} \left[C_0 + 2 \sum_{m=1}^\infty \cos (m \varphi)\,
 C_m \right]\, .
\end{eqnarray}
The hybrid high-energy and collinear factorization provides us with a general formula for the $C_m$ coefficients which holds at the NLO perturbative order and embodies the resummation of high-energy logarithms up to NLL accuracy. Working in the $\MSb$ renormalization scheme, one has~\cite{Caporale:2012ih}
\[
 C_m^{\rm NLL} = 
 \hspace{-0.15cm}
 \int_{q_1^{\rm inf}}^{q_1^{\rm sup}} \hspace{-0.35cm} \drv |\bm{q}|_1
 \int_{q_2^{\rm inf}}^{q_2^{\rm sup}} \hspace{-0.35cm} \drv |\bm{q}|_2
 \int_{y_1^{\rm inf}}^{y_1^{\rm sup}} \hspace{-0.35cm} \drv y_1
 \int_{y_2^{\rm inf}}^{y_2^{\rm sup}} \hspace{-0.35cm} \drv y_2
 \, \delta(y_1 - y_2 - \DY)
\]
\[
 \hspace{-0.33cm}
 \times \; \frac{e^{\DY}}{s} \hspace{-0.05cm} \int_{-\infty}^{+\infty} \hspace{-0.20cm} \drv \nu \, e^{{\DY} \bar \alpha_s\chi^{\rm NLO}(m,\nu)}
 \, \alpha_s^2(\mu_R)
 \bigg[ 
 \bar \alpha_s^2 \frac{\beta_0 \DY}{4 N_c}\chi(m,\nu)f(\nu)
\]
\begin{eqnarray}
\label{Cn_NLL_MSb}
 \hspace{-0.78cm}
 + \,\,
 c_1^{\rm NLO}(m,\nu,|\bm{q}_1|, x_1)[c_2^{\rm NLO}(m,\nu,|\bm{q}_2|,x_2)]^*
 \bigg] \;,
\end{eqnarray}
with $\bar \alpha_s(\mu_R) \equiv \alpha_s(\mu_R) N_c/\pi$, $N_c$ the number of colors, and $\beta_0 = 11N_c/3 - 2 n_f/3$.
The BFKL kernel at the exponent in Eq.~\eqref{Cn_NLL_MSb} encodes the NLL resummation of energy logarithms
\begin{eqnarray}
 \label{chi}
 \chi^{\rm NLO}(m,\nu) = \chi(m,\nu) + \bar\alpha_s \hat \chi(m,\nu) \;,
\end{eqnarray}
where $\chi(m,\nu)$ are the LO BFKL eigenvalues
\begin{eqnarray}
\chi\left(m,\nu\right) = 2\left\{\psi\left(1\right)-{\rm Re} \left[\psi\left( (m + 1)/2 + i \nu \right)\right] \right\}
\label{chi_LO}
\end{eqnarray}
with $\psi(z) = \Gamma^\prime(z)/\Gamma(z)$.
The $\hat\chi(m,\nu)$ function in Eq.\eref{chi} is the NLO kernel correction
\begin{eqnarray}
\label{chi_NLO}
\hat \chi\left(m,\nu\right) &=& \bar\chi(m,\nu)+\frac{\beta_0}{8 N_c}\chi(m,\nu)
\\ \nonumber &\times& \,
\left\{-\chi(m,\nu)+10/3+2\ln\left[\left(\mu_R^2/|\bm{q}_1| |\bm{q}_2|\right)\right]\right\}
\end{eqnarray}
with the characteristic function  $\bar\chi(m,\nu)$ calculated in~\cite{Kotikov:2000pm}. The two expressions
\begin{eqnarray}
\label{IFs}
c_{1,2}^{\rm NLO}(m,\nu,|\bm{q}|,x) =
c_{1,2} +
\alpha_s(\mu_R) \, \hat c_{1,2}
\end{eqnarray}
stand for the NLO impact factors.
We depict the emissions of heavy hadrons ($B_c^{*}$ and ${\cal H}_b$) \emph{via} the impact factor calculated in~\cite{Ivanov:2012iv}. Although being designed to study light hadrons, the use of this impact factor is valid in our VFNS approach, provided that transverse-momenta windows imposed for hadrons' detection are much larger than heavy-quark thresholds for DGLAP evolution.
The LO hadron impact factor takes the form of a collinear convolution
\begin{eqnarray}
\nonumber
c_h(m,\nu,|\bm{q}|,x) = \rho_c \, |\bm{q}|^{2i\nu-1}\int_{x}^1 \drv \xi / \xi \; \tilde{x}^{1-2i\nu} 
\nonumber \\
\label{LOHIF}
 \times \, \Big[\tau_c f_g(\xi)D_g^h\left(\tilde{x}\right)
 +\sum_{s=q,\bar q}f_s(\xi)D_s^h\left(\tilde{x}\right)\Big] \;,
\end{eqnarray}
where $\tilde{x} = x/\xi$, $\rho_c = 2 \sqrt{C_F/C_A}$, and $\tau_c = C_A/C_F$, with $C_F = (N_c^2-1)/(2N_c)$ and $C_A \equiv N_c$ the Casimir factors related to a gluon emission from a quark and from a gluon, respectively. Then $f_{s}\left(x, \mu_F \right)$ is the PDF for the parton $s$ extracted from the incoming proton and $D^h_{s}\left(x/\beta, \mu_F \right)$ denote the collinear FF for the parton $s$ fragmenting to the tagged hadron, $h \equiv \{B_c^{*}, {\cal H}_b\}$.
The NLO hadron impact factor was calculated in~\cite{Ivanov:2012iv}.
The LO light-jet impact factor is
\begin{eqnarray}
 \label{LOJIF}
 \hspace{-0.09cm}
 c_J(m,\nu,|\bm{q}|,x) = \rho_c
 |\bm{q}|^{2i\nu-1}\,\hspace{-0.05cm}\Big(\tau_c f_g(x)
 +\hspace{-0.15cm}\sum_{s=q,\bar q}\hspace{-0.10cm}f_s(x)\Big) \;.
\end{eqnarray}
The formula for the NLO jet impact factor can be got by combining Eq.~(36) of~\cite{Caporale:2012ih} with Eqs.~(4.19)-(4.20) of~\cite{Colferai:2015zfa}.
It is based on a small-cone algorithm~\cite{Ivanov:2012ms} with the jet-cone radius fixed at ${\cal R}_J = 0.5$, as usually employed in recent studies at CMS~\cite{Khachatryan:2016udy}.
The $f(\nu)$ function in Eq.~\eqref{Cn_NLL_MSb} reads
\begin{eqnarray}
 f(\nu) = \frac{i}{2} \left[ \frac{\drv}{\drv \nu} \ln(c_1/c_2^*) + \ln\left(|\bm{q}_1| |\bm{q}_2|\right) \right] \;.
\label{fnu}
\end{eqnarray}

In order to compare our NLL predictions with a reference fixed-order calculation, we truncate the expansion of azimuthal coefficients in Eq.~(\ref{Cn_NLL_MSb}) up to ${\cal O}(\alpha_s^3)$. In this way we obtain an effective high-energy fixed-order (HE-NLO) expression that can be handily employed in our phenomenological study.
It encodes leading-power asymptotic signal present in a pure NLO DGLAP calculation, and does not contain those factors which are suppressed by inverse powers of the partonic center-of-mass energy.
The $\MSb$ formula for our $C_m$ coefficients in the HE-NLO limit is
\[
 \CmHENLO = 
 \hspace{-0.15cm}
 \int_{q_1^{\rm inf}}^{q_1^{\rm sup}} \hspace{-0.35cm} \drv |\bm{q}|_1
 \int_{q_2^{\rm inf}}^{q_2^{\rm sup}} \hspace{-0.35cm} \drv |\bm{q}|_2
 \int_{y_1^{\rm inf}}^{y_1^{\rm sup}} \hspace{-0.35cm} \drv y_1
 \int_{y_2^{\rm inf}}^{y_2^{\rm sup}} \hspace{-0.35cm} \drv y_2
 \, \delta(y_1 - y_2 - \DY)
\]
\[
 \hspace{-0.33cm}
 \times \; \frac{e^{\DY}}{s} \hspace{-0.05cm} \int_{-\infty}^{+\infty} \hspace{-0.20cm} \drv \nu
 \, \alpha_s^2(\mu_R)
 \big[ 
 \bar \alpha_s(\mu_R) \DY \chi(m,\nu) 
\]
\begin{eqnarray}
\label{Cn_HENLO_MSbar}
  \hspace{-0.34cm}
 + \; c_1^{\rm NLO}(m,\nu,|\bm{q}_1|, x_1)[c_2^{\rm NLO}(m,\nu,|\bm{q}_2|,x_2)]^* \big] \,,
\end{eqnarray}
where the exponentiated BFKL kernel underwent an expansion up to ${\cal O}(\alpha_s)$.
Finally, by neglecting all NLO terms in Eqs.~\eqref{chi_NLO} and~\eqref{IFs}, we get a LO/LL limit for $C_m$ coefficients.

We set renormalization and factorization scales at the \emph{natural} energies of the process. Thus we fix $\mu_F = \mu_R = \mu_N \equiv m_{1 \perp} + m_{2 \perp}$, with $m_{i \perp} = \sqrt{ m_i^2 + \bm{q}_i^2}$ the transverse mass of the observed $i$ particle. Particle masses used in our phenomenology are $m_{B_c} = 6.275$ GeV and $m_{{\cal H}_b} \equiv m_{\Lambda_b} = 5.62$ GeV. Since in our treatment jet mass corrections are not considered, we can safely set $m_{2 \perp} \equiv |\bm{q}_2|$ in jet emissions.

\subsection{Collinear inputs}
\label{ssec:PDFs_FFs}

\begin{figure*}[!t]
\centering

   \includegraphics[scale=0.51,clip]{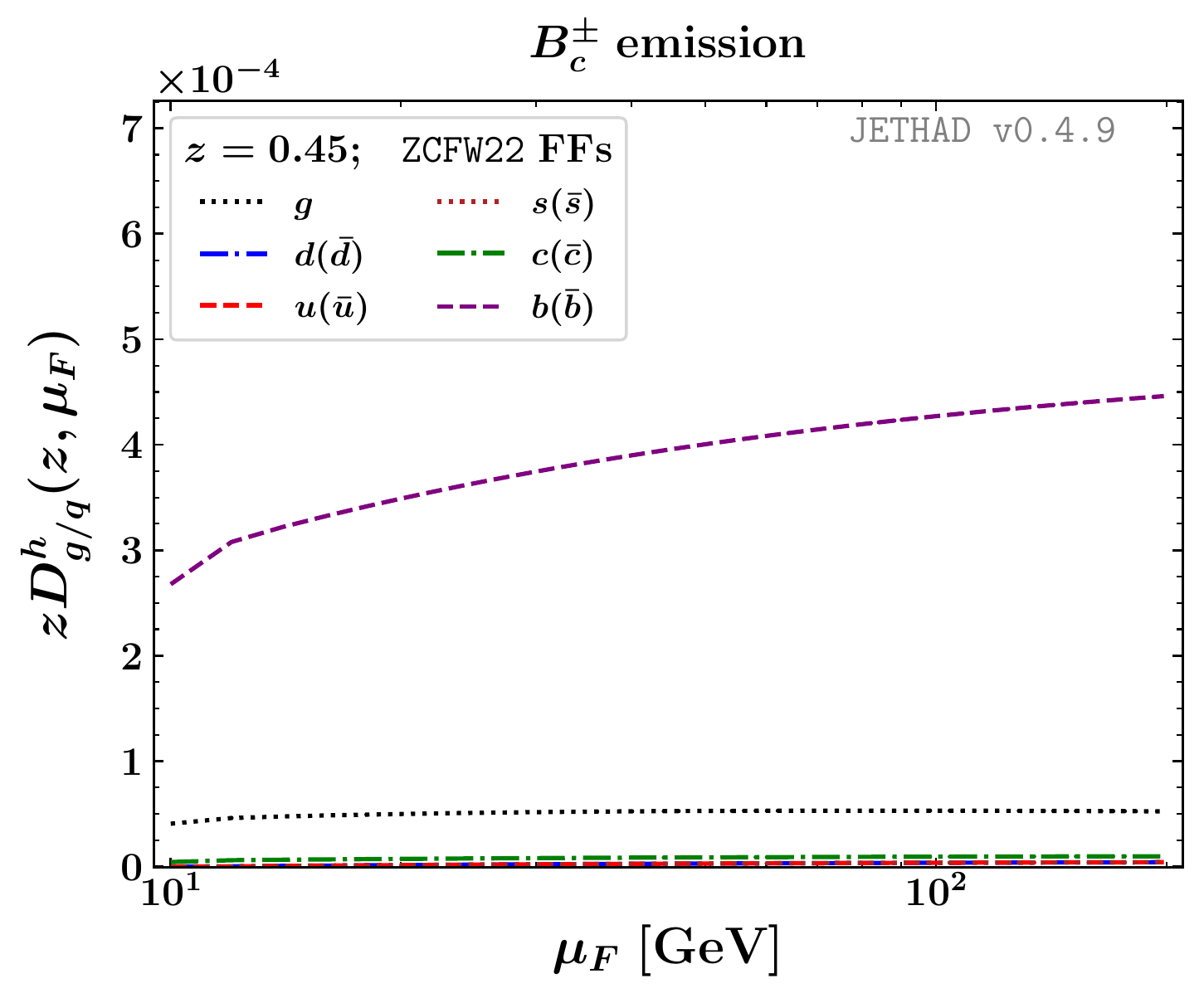}
   \hspace{0.10cm}
   \includegraphics[scale=0.51,clip]{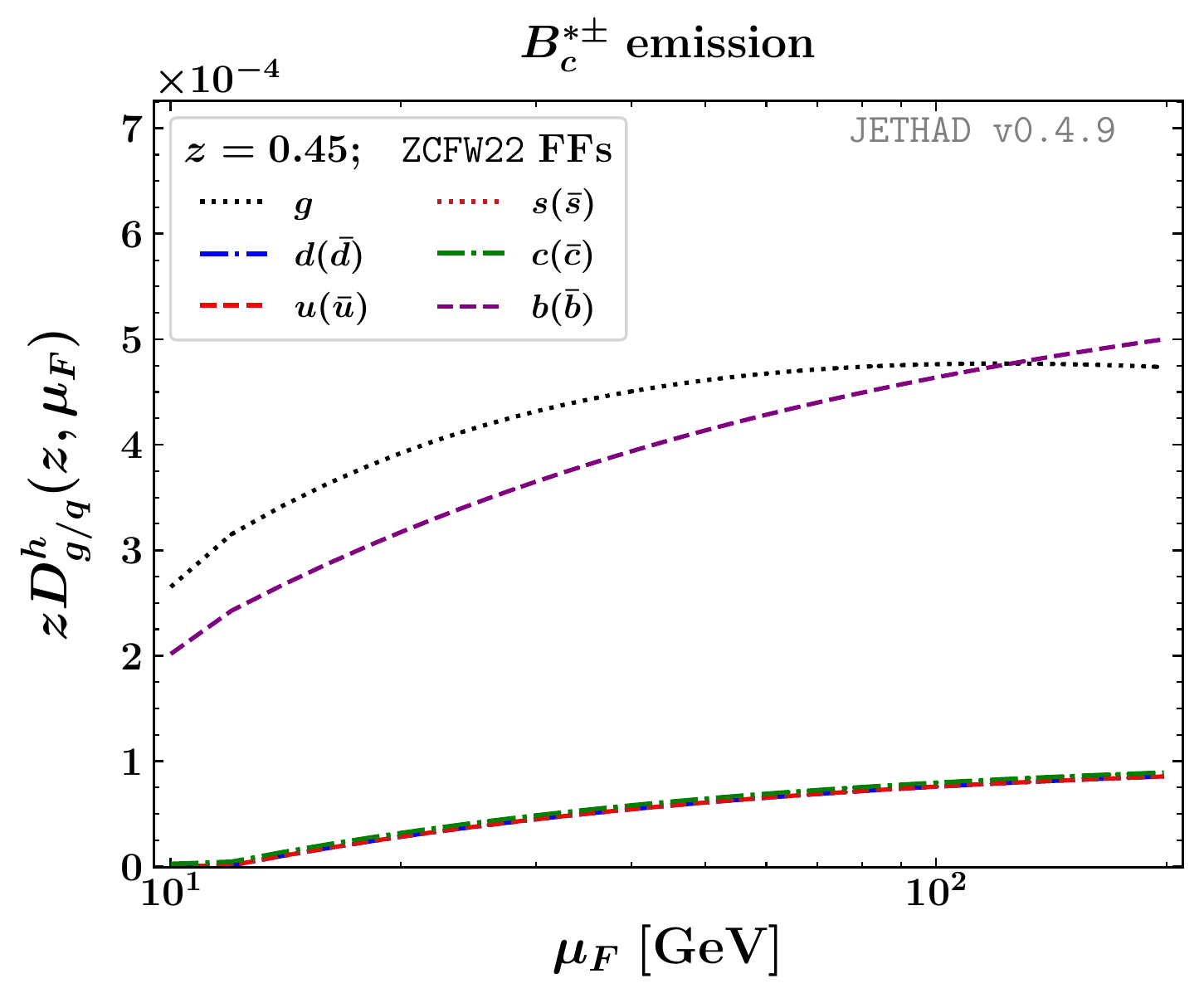}

\caption{{\tt ZCFW22} NLO FFs for $B_c$ (left) and $B_c^*$ (right)
production, as functions of $\mu_F$ and for $z \equiv \langle z \rangle \simeq 0.45$.}
\label{fig:FFs}
\end{figure*}

We employ NLO collinear PDFs extracted by the {\tt MMHT14} Group~\cite{Harland-Lang:2014zoa}.
In order to describe $B_c^{(*)}$ emissions, we have built a DGLAP-evolved FF set relying on corresponding NRQCD inputs. We combined the initial-scale heavy-quark ($c$ or $b$) NLO FFs calculated in~\cite{Zheng:2019gnb} with the gluon NLO FFs obtained in~\cite{Zheng:2021sdo}, and we performed the DGLAP evolution through the {\tt APFEL} package~\cite{Bertone:2013vaa}. This allowed us to generate a novel {\tt LHAPDF}~\cite{Buckley:2014ana} FF set, ready for phenomenology, which we called {\tt ZCFW22} set.
${\cal H}_b$ hadrons are depicted \emph{via} {\tt KKSS07} NLO FFs~\cite{Kniehl:2011bk,Kramer:2018vde}.

The FF characteristic pattern shown in Fig.~\ref{fig:FFs} for a momentum fraction $z$ which approximately corresponds to its average value, $\langle z \rangle \simeq 0.45$, gives us a clear indication that strong stabilizing effects are expected when $B_c^{(*)}$ mesons are emitted at high energies.
Indeed, as pointed out in~\cite{Celiberto:2021dzy,Celiberto:2021fdp,Celiberto:2022dyf}, in our hybrid factorization a key role is played by the heavy-hadron gluon FF. Its impact on the cross section is enhanced by the convolution with the gluon PDF in the LO hadron impact factor (Eq.~\eqref{LOHIF}), which prevails also over the non-diagonal $gq$ channel opened at NLO.
The smooth-behaved, non-decreasing with $\mu_F$ gluon FFs in both panels of Fig.~\ref{fig:FFs} compensate the lowering with $\mu_R$ of the running coupling in the
exponentiated kernel and the impact factors, thus generating the stability observed in heavy-flavor high-energy cross sections (see, \emph{e.g.}, Appendix of~\cite{Celiberto:2021fdp} for details).

In FFs depicting $\Lambda_c$~\cite{Kniehl:2020szu} and ${\cal H}_b$~\cite{Kniehl:2011bk} emissions, as well as in the NRQCD-based {\tt ZCW19} set for $\JPsi$ and $\Yps$~\cite{Zheng:2019dfk,Celiberto:2022dyf}, initial-scale light-parton FFs are set to zero, so that gluons and light quarks are generated by DGLAP evolution only.
Conversely, as mentioned, our {\tt ZCFW22} FFs for $B_c^{(*)}$ mesons contain a non-zero initial-scale gluon input from NRQCD~\cite{Zheng:2021sdo}. This provides us with a decisive evidence that the \emph{natural stabilization} of the high-energy resummation is an \emph{intrinsic property} of heavy-flavor emissions, which becomes manifest whenever a heavy-hadron species is detected, regardless of \emph{Ans\"atze} made on corresponding FFs.

\section{Phenomenology}
\label{sec:pheno}

Analyses of our observables were performed \emph{via} the numerical implementation of NLO/NLL distributions provided by the {\tt JETHAD} multi-modular interface~\cite{Celiberto:2020wpk}.
We studied the sensitivity of our observables by varying $\mu_R$ and $\mu_F$ scales around their natural values, up to a factor going from 1/2 to two.
The $C_{\mu}$ parameter entering figures stands for the ratio $C_\mu = \mu_{R,F}/\mu_N$. Uncertainty bands in our plots encode the overall effect of scale variation and phase-space multi-dimensional computation. The latter was constantly kept below 1\% by {\tt JETHAD} integration algorithms.

$B_c^{(*)}$ as well as ${\cal H}_b$ particles are detected inside rapidity acceptances of the CMS barrel, namely from $-2.4$ and $2.4$. The light jet can be reconstructed also by CMS endcaps~\cite{Khachatryan:2016udy}, thus having $|y_2| < 4.7$.
To respect the validity range of the fragmentation approximation~\cite{Kolodziej:1995nv,Artoisenet:2007xi}, transverse momenta of charmed $B$-mesons lie in the range $60 < |\bm{q}_1|/{\rm GeV} < 120$.
An asymmetric choice for the $\cal P$ particle transverse momentum, useful to better disengage high-energy resummation signals from the fixed-order background~\cite{Ducloue:2013bva,Celiberto:2015yba,Celiberto:2020wpk}, is realized by letting $30 < |\bm{q}_2|/{\rm GeV} < 120$ (${\cal H}_b$) and $50 < |\bm{q}_2|/{\rm GeV} < 120$ (jet).
Our selection for $|\bm{q}_{1,2}|$ ranges is in line with the spirit of the VFNS approach, whose validity holds when energy scales are sufficiently larger than thresholds for the DGLAP evolution given by $c$- and $b$-quark masses.
The center-of-mass energy is $\sqrt{s} = 14$ TeV.

\subsection{Rapidity distribution}
\label{ssec:DY}

\begin{figure*}[!t]
\centering

   \includegraphics[scale=0.51,clip]{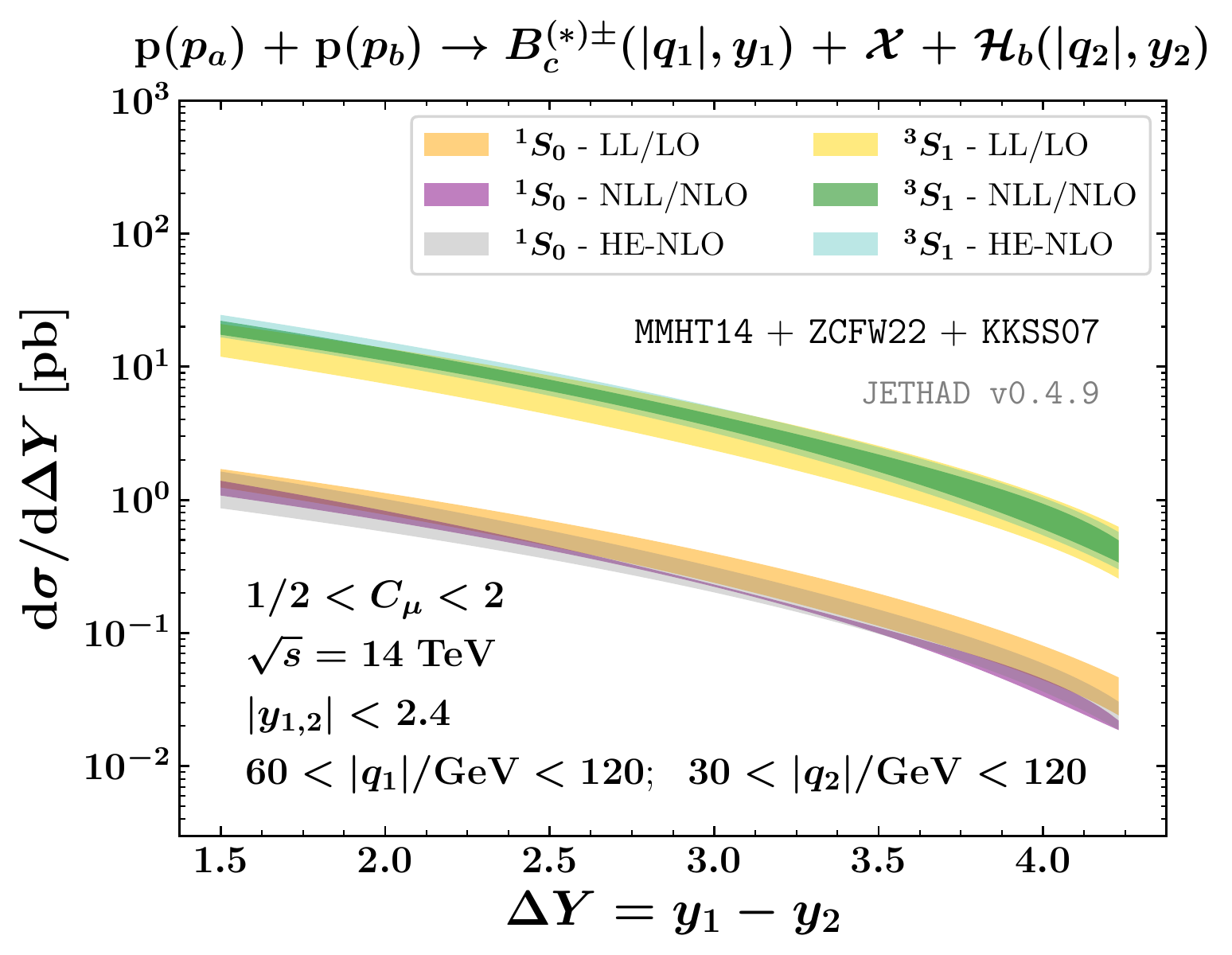}
   \hspace{0.10cm}
   \includegraphics[scale=0.51,clip]{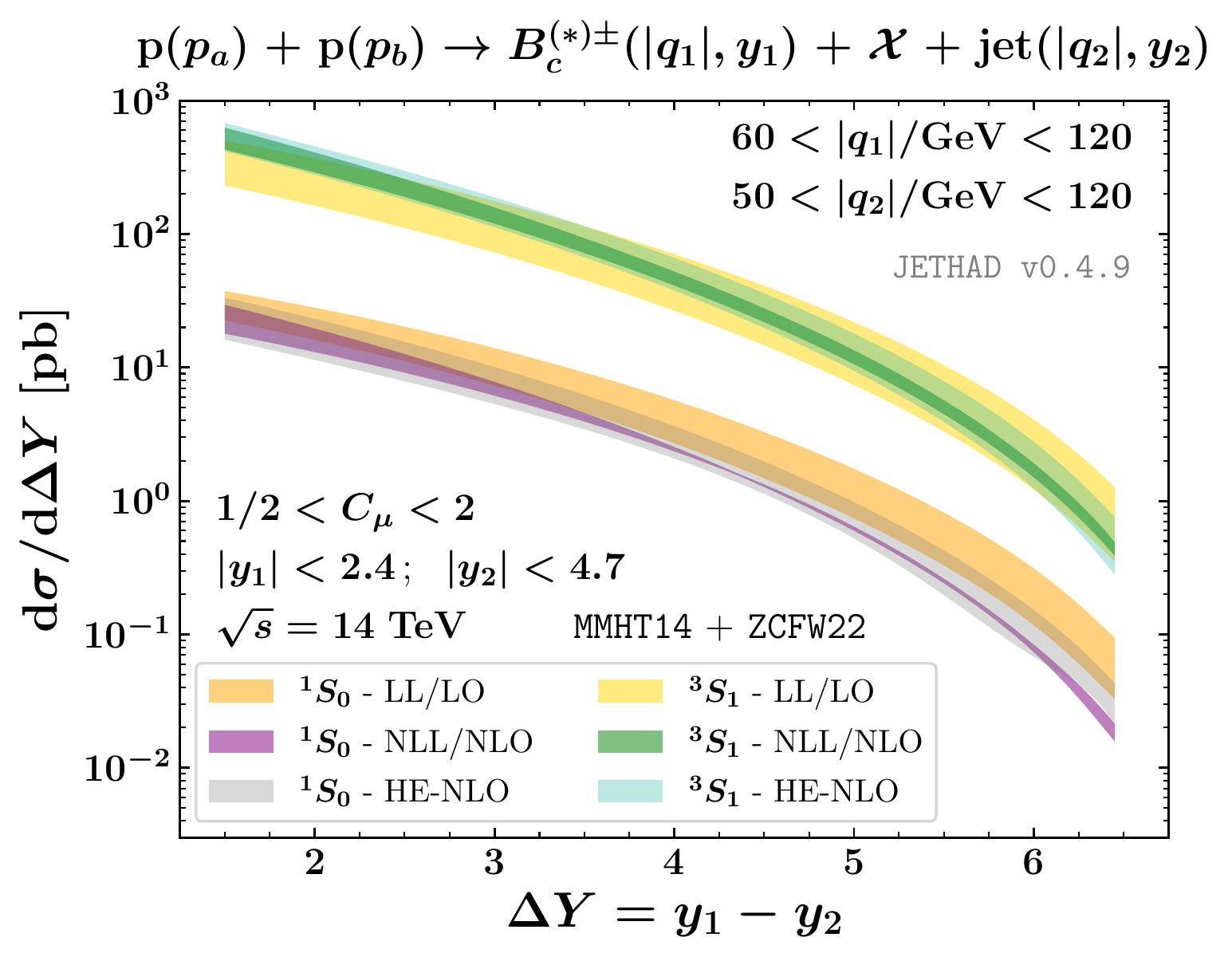}

\caption{Rapidity distribution for $B_c^{(*)} + {\cal H}_b$ (left) and $B_c^{(*)} + {\rm jet}$ (right) production at $\sqrt{s} = 14$~TeV. Shaded bands exhibit the combined uncertainty of scale variation and phase-space multi-dimensional integration.}
\label{fig:rapidity_distribution}
\end{figure*}

We consider the cross section differential in the final-state rapidity interval, $\DY$, obtained by integrating the r.h.s. of Eq.~\eqref{dsigma_Fourier} over the azimuthal angle distance, $\varphi$. It genuinely corresponds to the first azimuthal coefficient, $C_0$.
Panels of Fig.\tref{fig:rapidity_distribution} show the $\DY$-behavior of $C_0$ for $B_c^{(*)} + {\cal H}_b$ (left) and $B_c^{(*)} + {\rm jet}$ (right) reactions. The common trend is a downfall at large $\DY$. It comes out as the net result of two competing features: high-energy resummed hard factors increase with $\DY$ and thus with energy, as predicted by BFKL, but their collinear convolution with PDFs and FFs in the impact factors heavily quenches that growth.

We observe that NLL/NLO bands for emissions of $B_c(^3S_1)$ resonances are constantly smaller and almost entirely nested inside LL/LO ones. This is a clear signal that the resummed series has reached a fair stability under higher-order corrections and energy-scale variations. It represents a significant and required step in the path toward precision studies of our reactions at high energies. The stabilization pattern is milder in the $B_c(^1S_0)$ case, and the main reason is that the $B_c$ {\tt ZCFW22} gluon FF has a non-decreasing trend with $\mu_F$ (Fig.\tref{fig:FFs}, left panel), while the corresponding $B_c^{(*)}$ function visibly increases with $\mu_F$ (right panel).

The heavier is the growth of the heavy-flavor gluon FF, the stronger is the stabilizing power coming from the balance of evolution effects on renormalization and factorization scales. The larger size of the $B_c$ gluon FF with respect to the $B_c^{(*)}$ ones determines the hierarchy of predictions for $C_0$.
We note that $B_c^{(*)} + {\cal H}_b$ distributions are more stable than $B_c^{(*)} + {\rm jet}$ ones. In the first case {\tt KKSS07} FFs depicting ${\cal H}_b$ emissions act as a further stabilizer in a way that cannot be achieved by light-jet detections. In all cases bands for fixed-order HE-NLO predictions are substantially overlapped with NLL/NLO and LL/LO ones, and sometimes they stay in between. Therefore, at this stage a search for a clear discrimination between BFKL and the fixed-order background still remains open (see Section\tref{ssec:phi}).

\subsection{Azimuthal distribution}
\label{ssec:phi}

We study the azimuthal distribution, namely the normalized cross section differential in $\varphi$ and $\DY$
\begin{eqnarray}
 \label{azimuthal_distribution}
 \frac{1}{\sigma} \frac{\drv \sigma}{\drv \varphi \, \drv \DY} =
 \frac{1}{\pi} \left[\frac{1}{2} + \sum_{m=1}^\infty \cos (m \varphi)\,
 \langle \cos(m \varphi) \rangle \right]\, ,
\end{eqnarray}
where the mean values of cosines are the so called azimuthal-correlation moments, $\langle \cos(m \varphi) \rangle \equiv C_m/C_0$.
It represents one of the most solid observables where to hunt for high-energy imprints. Indeed, it collects signals coming from all azimuthal modes, and not just from $C_0$ or from a single ratio $C_m/C_0$. Furthermore, being differential in $\varphi$, it eases the comparison with experimental data, whose kinematic spectrum is limited by the fact that detectors generally do not cover the full $(2\pi)$ range.

\begin{figure*}[!t]
\centering


   \includegraphics[scale=0.51,clip]{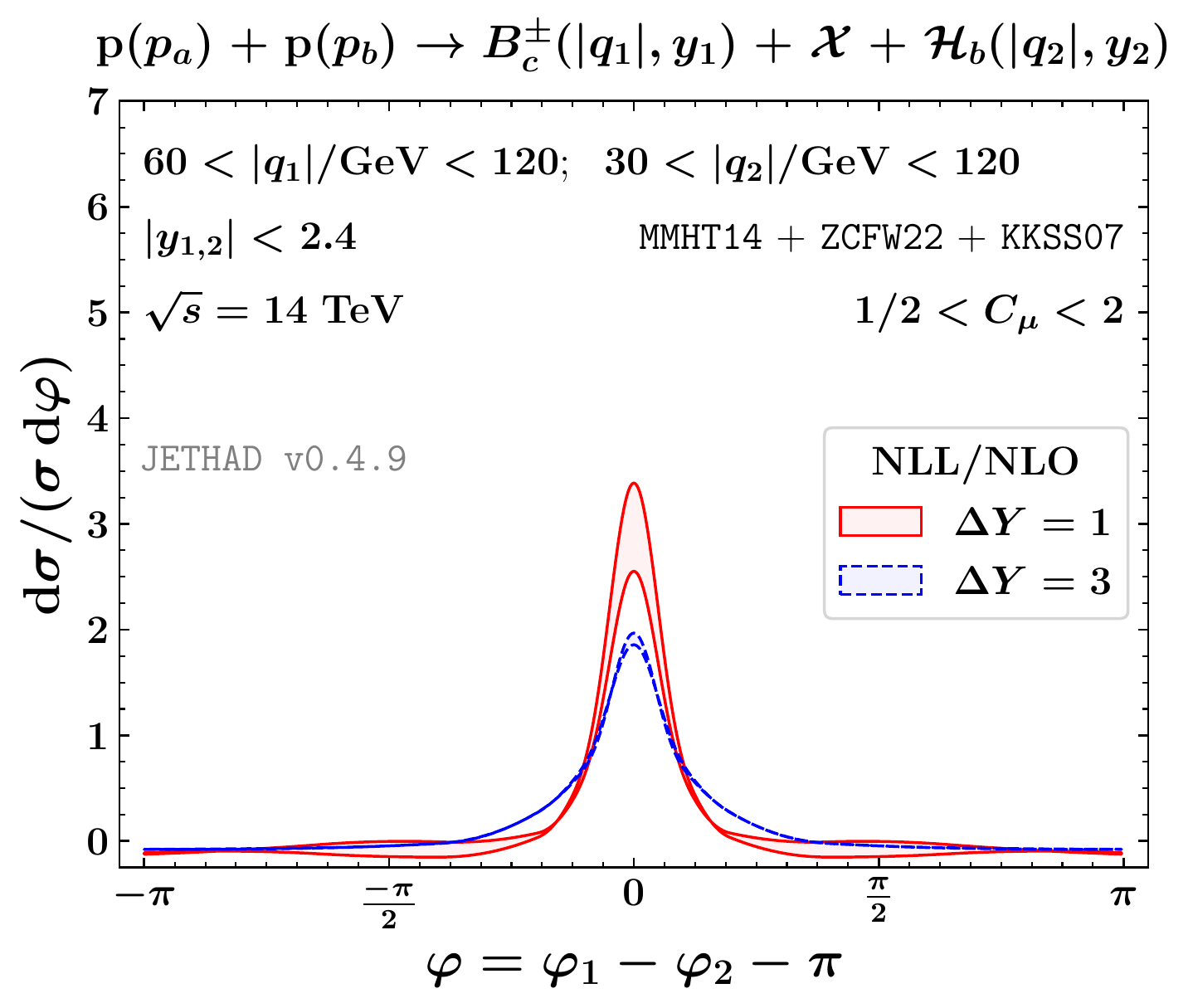}
   \hspace{0.10cm}
   \includegraphics[scale=0.51,clip]{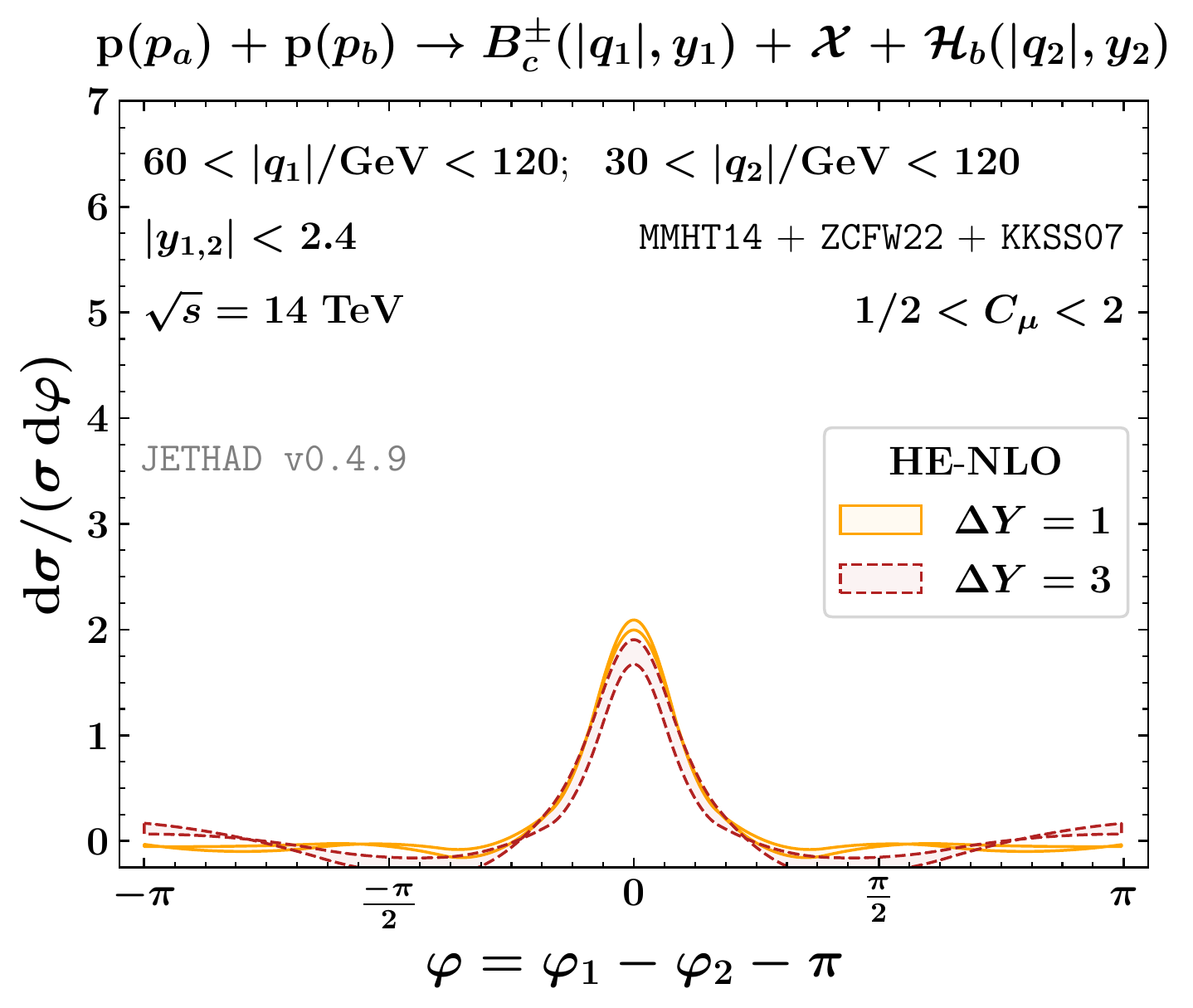}

   \includegraphics[scale=0.51,clip]{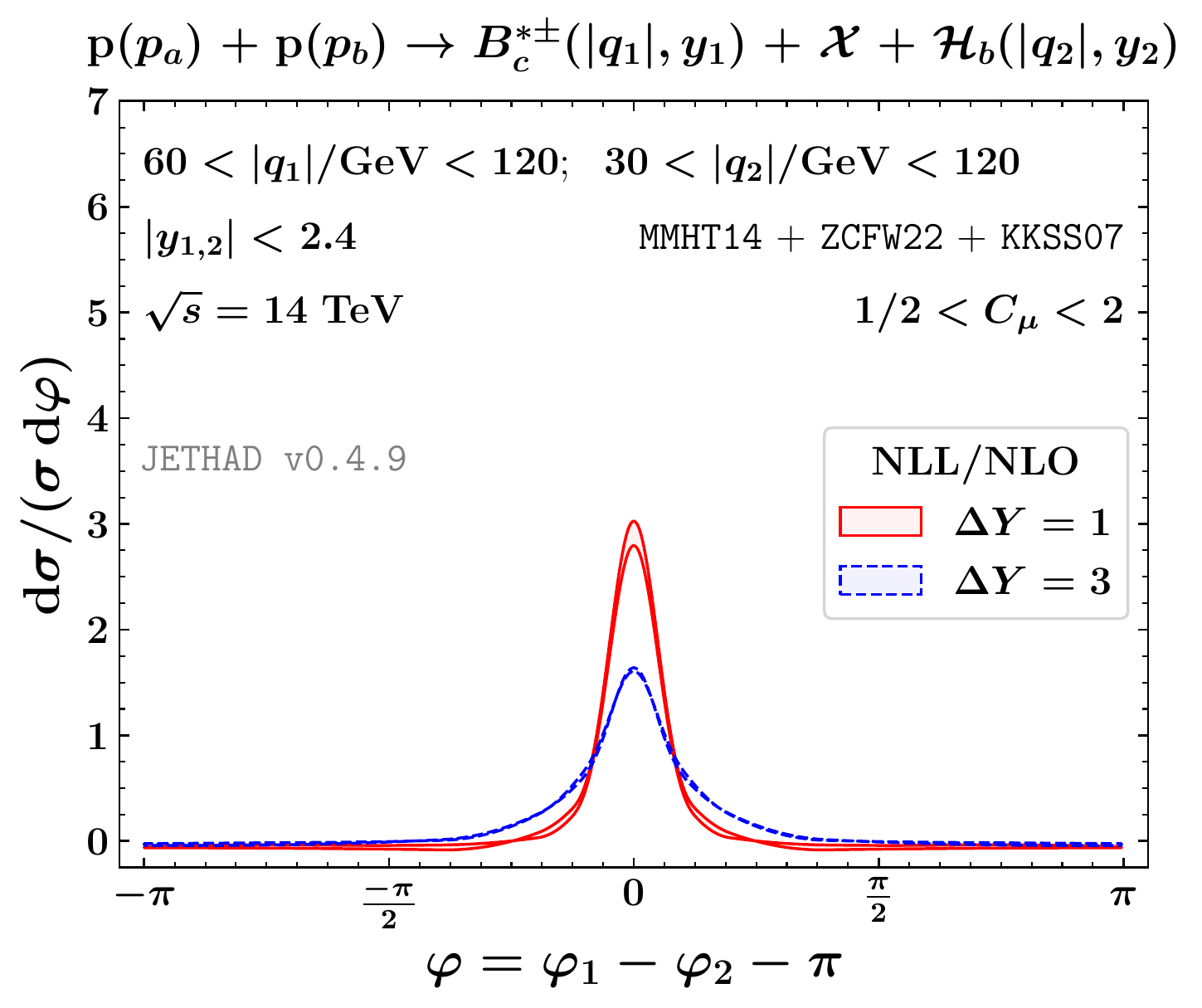}
   \hspace{0.10cm}
   \includegraphics[scale=0.51,clip]{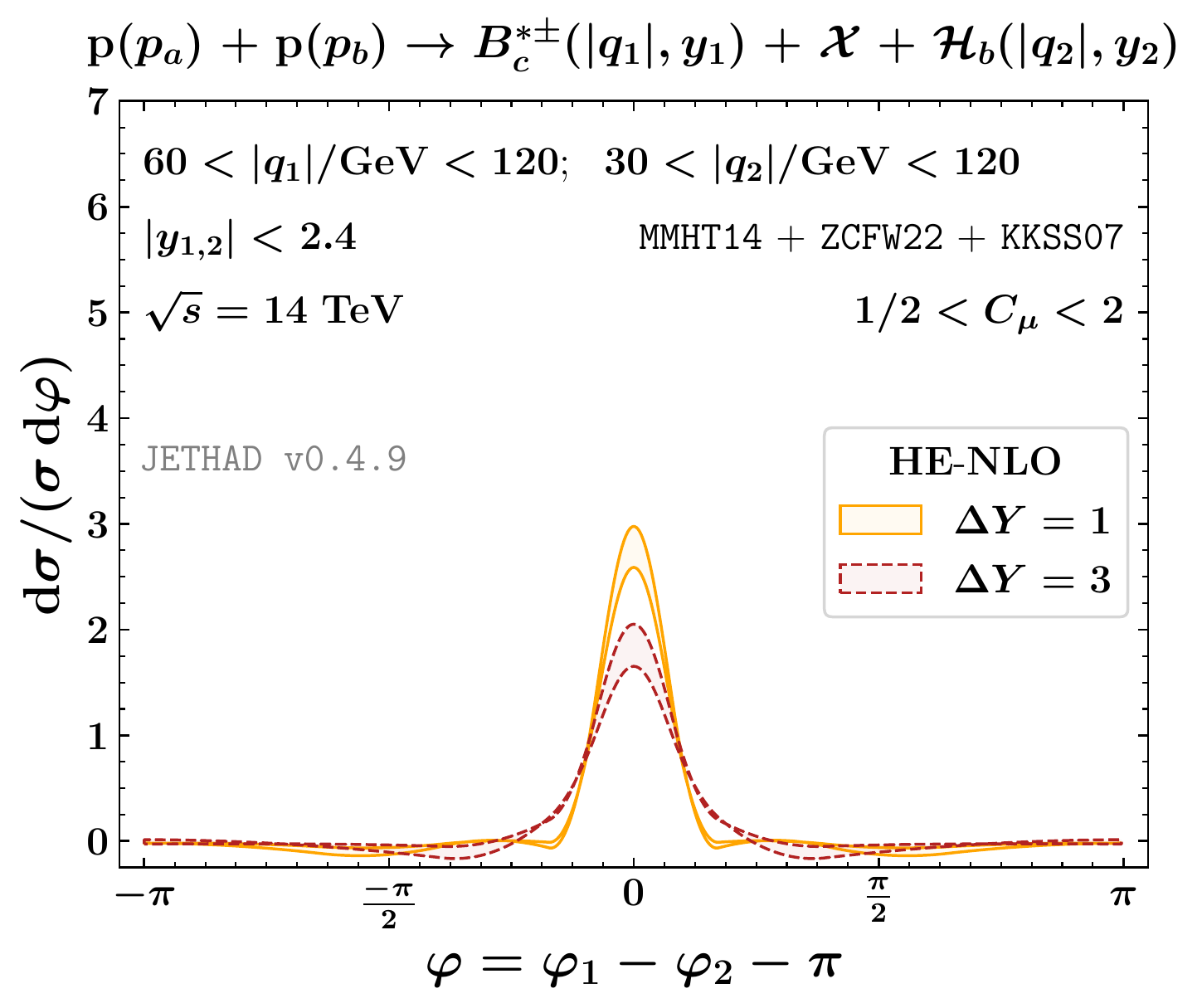}

\caption{Azimuthal distribution for $B_c^{(*)} + {\cal H}_b$ final states at $\sqrt{s} = 14$~TeV. Shaded bands exhibit the combined uncertainty of scale variation and phase-space multi-dimensional integration.}
\label{fig:azimuthal_distribution_Bb}
\end{figure*}

\begin{figure*}[!t]
\centering


   \includegraphics[scale=0.51,clip]{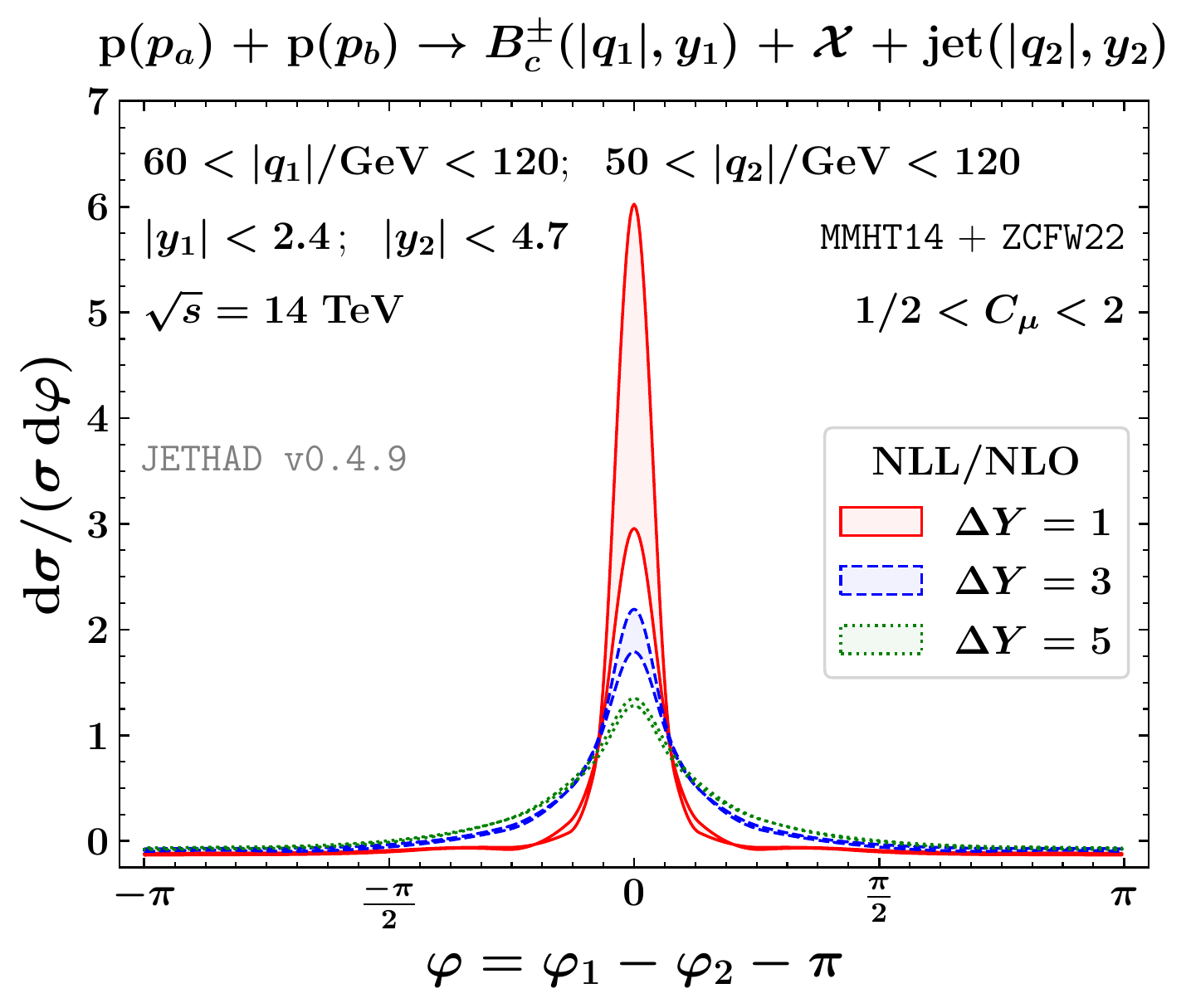}
   \hspace{0.10cm}
   \includegraphics[scale=0.51,clip]{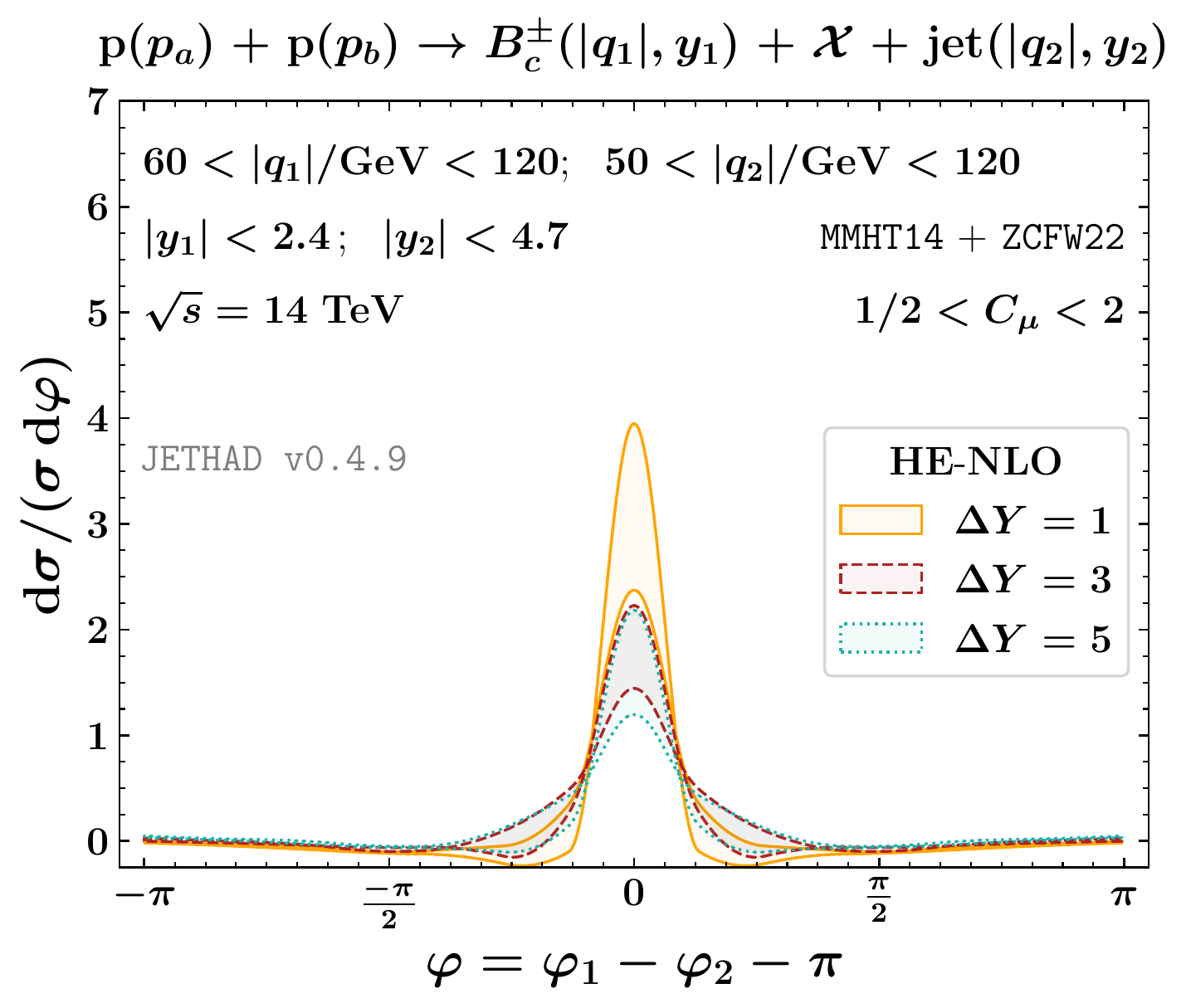}

   \includegraphics[scale=0.51,clip]{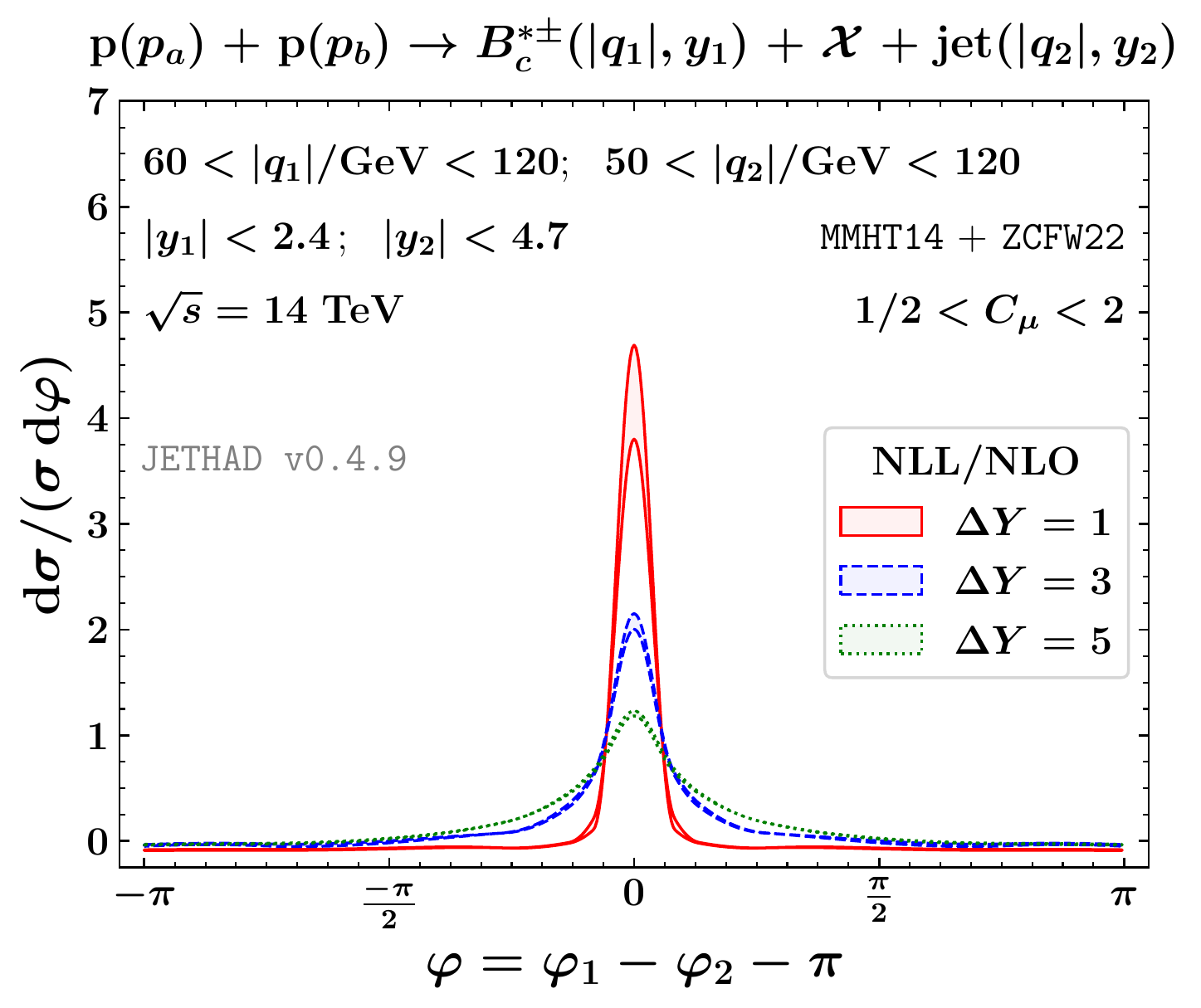}
   \hspace{0.10cm}
   \includegraphics[scale=0.51,clip]{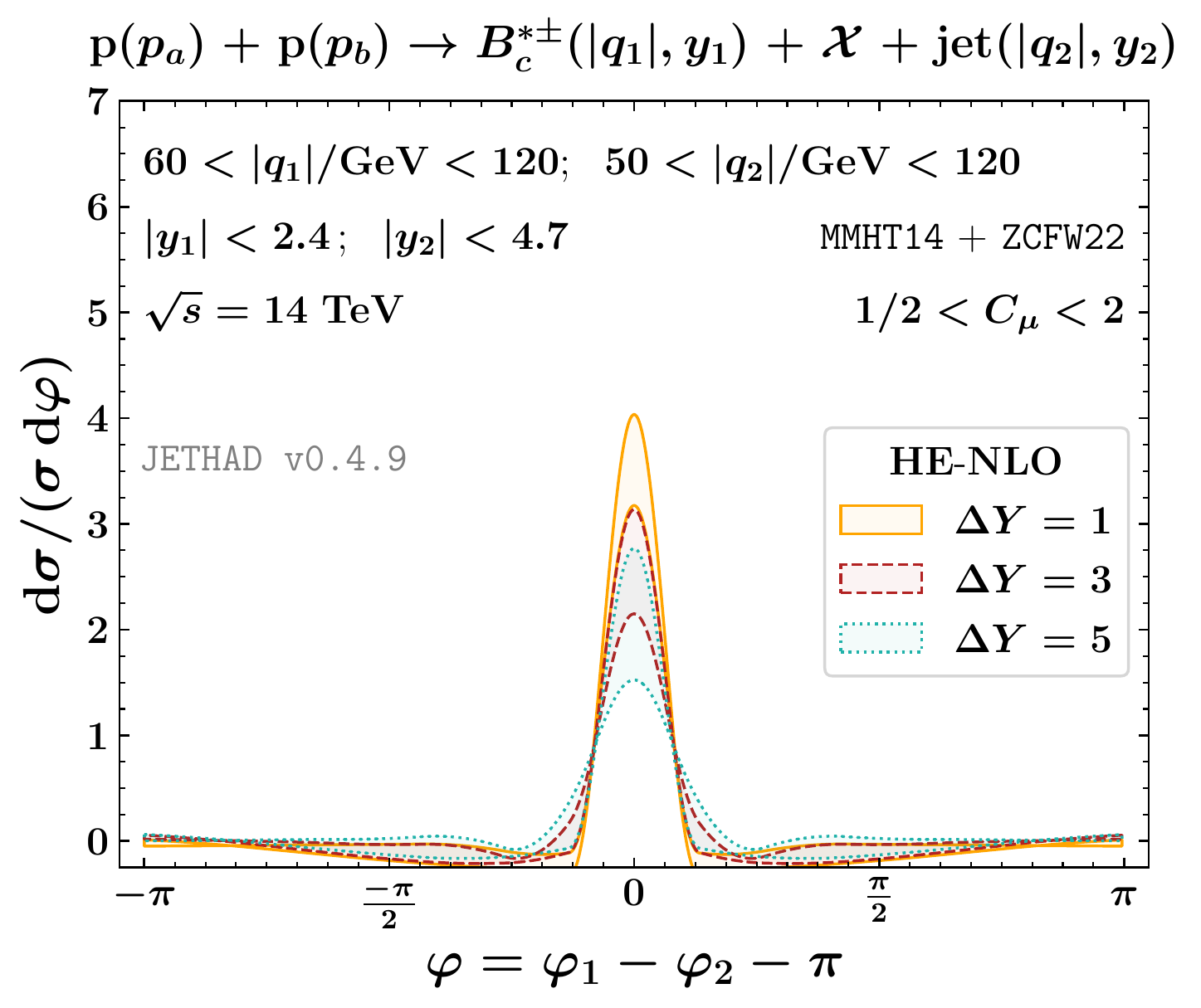}

\caption{Azimuthal distribution for $B_c^{(*)} + {\rm jet}$ final states at $\sqrt{s} = 14$~TeV. Shaded bands exhibit the combined uncertainty of scale variation and phase-space multi-dimensional integration.}
\label{fig:azimuthal_distribution_BJ}
\end{figure*}

Figures\tref{fig:azimuthal_distribution_Bb} and\tref{fig:azimuthal_distribution_BJ} contains predictions for azimuthal distributions at fixed values of $\DY$, for the $B_c^{(*)} + {\cal H}_b$ and $B_c^{(*)} + {\rm jet}$ channels, respectively. All the distributions exhibit a clear peak around $\varphi = 0$, namely when the two outgoing particles are emitted (almost) back-to-back. In the NLL/NLO case (left plots) the dominating peak is the one at the lowest value of the rapidity distance, $\DY = 1$. Then, when $\DY$ grows, the peak height visibly decreases and the distribution width broadens.
The found pattern tells us that high-energy dynamics has come into play in our kinematic regime. Indeed, as expected from BFKL, the weight of undetected-gluon radiation (the ${\cal X}$ system in Eq.\eref{process}) accounted for by the logarithmic resummation increases with $\DY$. This translates in a reduction of the rate of back-to-back events and, consequently, in a loss of correlation in the azimuthal plane.
Conversely, HE-NLO azimuthal distributions (right plots) are less sensitive to $\DY$-variations, with peaks less pronounced and uncertainty bands partially nested. This is a consequence of the truncation at ${\cal O}(\alpha_s^3)$ of the expansion of azimuthal coefficients in Eq.~(\ref{Cn_NLL_MSb}), which means that the number of secondary gluons radiated is fixed, independently of $\DY$.

We note that uncertainty bands are generally narrower when $B_c^{*}$ resonances are detected. It comes out as a further manifestation of the stabilizing power of heavy-flavor FFs which, as previously mentioned, is stronger when the gluon FF increases with $\mu_F$ in our typical ranges of the hadron's momentum fraction (see Fig.\eref{fig:FFs}). This in particular helps to get a better description of our distributions at lower values of the rapidity interval, say $\DY \gtrsim 1$, when the BFKL approach is stretched to its limits of applicability. All these features supports the statement that azimuthal distributions are excellent observables whereby observing and quantifying distinctive effects of the onset of high-energy QCD dynamics. They also offer us a fertile ground where BFKL-driven signals can be effectively disengaged from fixed-order results.

\section{Conclusions and Outlook}
\label{sec:conclusions}

We proposed the inclusive hadroproduction of a charmed $B$-meson accompanied by a non-charmed $b$-hadron or a light jet, as a novel tool to access the high-energy QCD dynamics. Their high rapidity distance and their large transverse momenta allowed us to describe rapidity and azimuthal distributions \emph{via} the hybrid high-energy and collinear factorization within the NLO/NLL.
Relying on the fragmentation approximation, we embodied in our formalism a new $B_c^{(*)}$ collinear FF set, named {\tt ZCFW22}. It was built \emph{via} the DGLAP evolution of NRQCD inputs for the fragmentation of gluon, $c$- and $b$-quarks into $B_c(^1S_0)$ and $B_c(^3S_1)$ states at NLO.

The fair stability exhibited by our distributions under higher-order corrections and energy-scale variations provides us with a further and stronger evidence that heavy-flavored emissions act as \emph{natural stabilizers} of the high-energy resummation.
The stabilization mechanism is encoded in the peculiar behavior of the heavy-hadron gluon FF, and it comes out as a general feature shared by all the heavy-flavored species investigated so far, heavy-light hadrons~\cite{Celiberto:2021dzy,Celiberto:2021fdp,Celiberto:2022zdg}, vector quarkonia~\cite{Celiberto:2022dyf}, and now $B_c^{(*)}$ mesons.
We believe that the favorable statistics of our cross sections motivates further analyses on the high-energy spectrum of QCD \emph{via} heavy flavor at the (high-luminosity) LHC~\cite{Chapon:2020heu,Amoroso:2022eow} and in ultra-forward directions of rapidity, as the ones accessible at the forthcoming Forward Physics Facility~(FPF)~\cite{Anchordoqui:2021ghd,Feng:2022inv,Celiberto:2022rfj}.

Our plan for future extensions of studies presented in this Letter is twofold.
First, we will investigate high-energy emissions of charmed $B$-mesons in lower transverse-momentum regimes, where short-distance mechanisms dominate over fragmentation. This will help us to: ($i$) shed light on $B_c^{(*)}$ production mechanisms, ($ii$) explore the interplay between the VFNS and other fixed-flavor approaches, and ($iii$) unveil the connection between the high-energy formalism and other resummations.
Then, we will assess the feasibility of studies on $B_c^{(*)}$ detections in forward directions at new-generation lepton~\cite{AlexanderAryshev:2022pkx} and lepton-hadron colliders~\cite{AbdulKhalek:2021gbh,Acosta:2022ejc}.

\section*{Acknowledgements}

The {\tt ZCFW22} FF set was built by partially making use of a native numerical code developed by Xu-Chang~Zheng, Chao-Hsi~Chang, Tai-Fu~Feng, and Xing-Gang~Wu~\cite{Zheng:2019gnb}. The author acknowledges permission on usage of that code.
The author is grateful to Alessandro~Papa and Mohammed M.A. Mohammed for a critical reading of this Letter and for encouragement.
The author would like to express his gratitude to Jean-Philippe~Lansberg, Hua-Sheng~Shao, and colleagues of the \emph{Quarkonia As Tools} Community for inspiring discussions on production mechanisms and phenomenology of heavy-flavored bound states.
This work was supported by the INFN/NINPHA project.
The author thanks the Universit\`a degli Studi di Pavia for the warm hospitality.

\bibliographystyle{elsarticle-num}

\bibliography{bibliography}

\end{document}